\documentclass[apj]{emulateapj}
\usepackage{apjfonts}
\usepackage{amsmath}
\usepackage{appendix}

\shorttitle{ACT Optical Richness - SZ Relation}
\shortauthors{Sehgal et al.}

\begin{document}

\title{The Atacama Cosmology Telescope: Relation Between Galaxy Cluster Optical Richness and Sunyaev-Zel'dovich Effect}

\author{
Neelima~Sehgal\altaffilmark{1},
Graeme~Addison\altaffilmark{2},
Nick~Battaglia\altaffilmark{3},
Elia~S.~Battistelli\altaffilmark{4,5},
J.~Richard~Bond\altaffilmark{6},
Sudeep~Das\altaffilmark{7},
Mark~J.~Devlin\altaffilmark{8},
Joanna~Dunkley\altaffilmark{2},
Rolando~D\"{u}nner\altaffilmark{9},
Megan~Gralla\altaffilmark{10},
Amir~Hajian\altaffilmark{6},
Mark~Halpern\altaffilmark{5},
Matthew~Hasselfield\altaffilmark{5},
Matt~Hilton\altaffilmark{11},
Adam~D.~Hincks\altaffilmark{6,12},
Ren\'ee~Hlozek\altaffilmark{1},
John~P.~Hughes\altaffilmark{13},
Arthur~Kosowsky\altaffilmark{14},
Yen-Ting~Lin\altaffilmark{15,16},
Thibaut~Louis\altaffilmark{2},
Tobias~A.~Marriage\altaffilmark{10},
Danica~Marsden\altaffilmark{17},
Felipe~Menanteau\altaffilmark{13},
Kavilan~Moodley\altaffilmark{18},
Michael~D.~Niemack\altaffilmark{19},
Lyman~A.~Page\altaffilmark{12},
Bruce~Partridge\altaffilmark{20},
Erik~D.~Reese\altaffilmark{8},
Blake~D.~Sherwin\altaffilmark{12},
Jon~Sievers\altaffilmark{6},
Crist\'obal~Sif\'on\altaffilmark{9},
David~N.~Spergel\altaffilmark{1},
Suzanne~T.~Staggs\altaffilmark{12},
Daniel~S.~Swetz\altaffilmark{19},
Eric~R.~Switzer\altaffilmark{6},
Ed~Wollack\altaffilmark{21}
}
\altaffiltext{1}{Department of Astrophysical Sciences, Peyton Hall,
Princeton University, Princeton, NJ USA 08544}
\altaffiltext{2}{Department of Astrophysics, Oxford University, Oxford,
UK OX1 3RH}
\altaffiltext{3}{Department of Physics, Carnegie Mellon University, Pittsburgh, PA 15213}
\altaffiltext{4}{Department of Physics, University of Rome ``La Sapienza'',
Piazzale Aldo Moro 5, I-00185 Rome, Italy}
\altaffiltext{5}{Department of Physics and Astronomy, University of
British Columbia, Vancouver, BC, Canada V6T 1Z4}
\altaffiltext{6}{Canadian Institute for Theoretical Astrophysics, University of
Toronto, Toronto, ON, Canada M5S 3H8}
\altaffiltext{7}{Berkeley Center for Cosmological Physics, LBL and
Department of Physics, University of California, Berkeley, CA, USA 94720}
\altaffiltext{8}{Department of Physics and Astronomy, University of
Pennsylvania, 209 South 33rd Street, Philadelphia, PA, USA 19104}
\altaffiltext{9}{Departamento de Astronom{\'{i}}a y Astrof{\'{i}}sica,
Facultad de F{\'{i}}sica, Pontificia Universidad Cat\'{o}lica de Chile,
Casilla 306, Santiago 22, Chile}
\altaffiltext{10}{Dept. of Physics and Astronomy, The Johns Hopkins University, 3400 N. Charles St., Baltimore, MD 21218-2686}
\altaffiltext{11}{Centre for Astronomy \& Particle Theory, School of Physics \& Astronomy,
University of Nottingham, University Park, Nottingham, NG7 2RD, UK}
\altaffiltext{12}{Joseph Henry Laboratories of Physics, Jadwin Hall,
Princeton University, Princeton, NJ, USA 08544}
\altaffiltext{13}{Department of Physics and Astronomy, Rutgers,
The State University of New Jersey, Piscataway, NJ USA 08854-8019}
\altaffiltext{14}{Department of Physics and Astronomy, University of Pittsburgh,
Pittsburgh, PA, USA 15260}
\altaffiltext{15}{Institute of Astronomy \& Astrophysics, Academia Sinica, Taipei, Taiwan}
\altaffiltext{16}{Institute for the Physics and Mathematics of the Universe,
The University of Tokyo, Kashiwa, Chiba 277-8568, Japan}
\altaffiltext{17}{Department of Physics, University of California Santa Barbara, CA 93106, USA}
\altaffiltext{18}{Astrophysics and Cosmology Research Unit, School of
Mathematical Sciences, University of KwaZulu-Natal, Durban, 4041,
South Africa}
\altaffiltext{19}{NIST Quantum Devices Group, 325
Broadway Mailcode 817.03, Boulder, CO, USA 80305}
\altaffiltext{20}{Department of Physics and Astronomy, Haverford College,
Haverford, PA, USA 19041}
\altaffiltext{21}{Code 553/665, NASA/Goddard Space Flight Center,
Greenbelt, MD, USA 20771}

\begin{abstract}
We present the measured Sunyaev-Zel'dovich (SZ) flux from 474 optically-selected MaxBCG clusters that fall within the Atacama Cosmology Telescope (ACT) Equatorial survey region.  The ACT Equatorial region used in this analysis covers 510 square degrees and overlaps Stripe 82 of the Sloan Digital Sky Survey.  We also present the measured SZ flux stacked on 52 X-ray-selected MCXC clusters that fall within the ACT Equatorial region and an ACT Southern survey region covering 455 square degrees.  We find that the measured SZ flux from the X-ray-selected clusters is consistent with expectations.  However, we find that the measured SZ flux from the optically-selected clusters is both significantly lower than expectations and lower than the recovered SZ flux measured by the {\em{{\it{Planck}}}} satellite.  Since we find a lower recovered SZ signal than {\em{{\it{Planck}}}},  we investigate the possibility that there is a significant offset between the optically-selected brightest cluster galaxies (BCGs) and the SZ centers, to which ACT is more sensitive due to its finer resolution. Such offsets can arise due to either an intrinsic physical separation between the BCG and the center of the gas concentration or from misidentification of the cluster BCG.  
We find that the entire discrepancy for both ACT and {\em{{\it{Planck}}}} can be explained by assuming that the BCGs are offset from the SZ maxima with a uniform random distribution between 0 and 1.5 Mpc. Such large offsets between gas peaks and BCGs for optically-selected cluster samples seem unlikely given that we find the physical separation between BCGs and X-ray peaks for an X-ray-selected subsample of MaxBCG clusters to have a much narrower distribution that peaks within 0.2 Mpc.  It is possible that other effects are lowering the ACT and Planck signals by the same amount, with offsets between BCGs and SZ peaks explaining the remaining difference between ACT and Planck measurements.   Several effects that can lower the SZ signal equally for both ACT and {\em{{\it{Planck}}}}, but not explain the difference in measured signals, include a larger percentage of false detections in the MaxBCG sample, a lower normalization of the mass-richness relation, radio or infrared galaxy contamination of the SZ flux, and a low intrinsic SZ signal. In the latter two cases, the effects would need to be preferentially more significant in the optically-selected MaxBCG sample than in the MCXC X-ray sample.  
\end{abstract}


\section{INTRODUCTION}\label{sec:intro}

Galaxy cluster properties may follow simple scaling laws reflecting their self-similarity.  This possibility has given credence to their use as cosmological probes.  Cosmological parameters have been obtained by X-ray, optical, and most recently SZ surveys of clusters \citep[e.g.,][]{Vikhlinin2009b, Mantz2010, Rozo2010, Sehgal2011,Benson2011}.  At the same time the scaling laws that feed into these parameter constraints continue to undergo scrutiny.  

Recent millimeter-wavelength data have opened a new window whereby these scaling relations can be robustly checked against SZ flux measurements.    The SZ cluster signal has been predicted to have a low-scatter correlation with cluster mass \citep[e.g.,][]{Motl2005, Nagai2006}.  If true, this would make SZ-detected clusters an excellent tracer of structure growth in the Universe \citep[e.g.,][]{Wang1998, Haiman2001, Holder2001, Carlstrom2002}.  As steps towards understanding the SZ-mass relation, several studies have shown that the SZ cluster signal correlates well with X-ray signals \citep[e.g.,][]{Bonamente2008, Bonamente2011, Andersson2011}, dynamically determined masses \citep{Sifon2012}, and weak-lensing determined masses \citep[e.g.,][]{Marrone2011}.  

In particular, the {\it{Planck}} satellite recently reported a good agreement between the measured and expected SZ-mass relation for a sample of X-ray-selected clusters \citep{Planck2011c,Planck2011a}.  However, a similar comparison for optically-selected clusters yielded an amplitude of SZ flux lower than expected by about a factor of two, with an even larger discrepancy for lower-mass clusters \citep{Planck2011b}.  An analysis by \citet{Draper2011} using data from the {\em{{\it{WMAP}}}} satellite found a similar result, however with larger uncertainty.  \citet{Hand2011}, using data from the Atacama Cosmology Telescope (ACT) \citep{Swetz2011} and stacking luminous red galaxies, also suggested a low SZ flux for optically-selected halos.  Among the possible explanations could be that either the SZ signal is not a robust tracer of galaxy clusters and groups or that optical selection techniques are somehow biased.  

Here, we investigate this discrepancy by stacking optically-selected clusters in millimeter-wavelength data from ACT that overlaps Stripe 82 of the Sloan Digital Sky Survey \citep{York2000}.  We also measure the SZ flux for X-ray-selected clusters as a consistency check.  Understanding these scaling relations will have important implications for cluster astrophysics as well as for their use in cosmological studies.  

This paper is organized as follows.  Section \ref{sec:dataSets} discusses the data sets used in this analysis.  Section \ref{sec:szMeas} describes the method used to measure cluster SZ flux.  Results are presented in Sections~\ref{sec:sims} through \ref{sec:planck} and discussed in Section \ref{sec:discuss}.  

\section{DATA SETS}\label{sec:dataSets}

Below we describe the catalogs of optically-selected and X-ray-selected clusters and the millimeter-wavelength data used to measure the cluster SZ fluxes.  We note that throughout this work $M_{500c}$ refers to the mass within $R_{500c}$, which is the radius within which the average density equals 500 times the critical density of the Universe at the cluster redshift.  Similarly, $M_{200m}$ gives the mass within $R_{200m}$, the radius within which the average density equals 200 times the mean matter density of the Universe at the cluster redshift.   A fiducial cosmology of $\Omega_m$ = 0.27, $\Omega_{\Lambda}$ = 0.73, and $h$ = 0.71 is also adopted \citep{Komatsu2011}, where $H(z) = H_0 E(z) = (h \times 100~{\rm{km~s^{-1}~Mpc^{-1}}}) E(z)$ and $E(z) = [\Omega_{m} (1+z)^3 + \Omega_{\Lambda}]^{1/2}$.

\subsection{The MaxBCG Optical Cluster Catalog}\label{maxbcg}

The MaxBCG Optical Cluster Catalog consists of 13,823 clusters selected from Data Release 5 (DR5) of the Sloan Digital Sky Survey \citep{Koester2007b, Koester2007a}.  The clusters were selected from a 7500 deg$^2$ area of sky using the observation that cluster galaxies tend to be the brightest galaxies at a given redshift, share a similar red color, and are spatially clustered.  The catalog consists of clusters that fall in the redshift range of $0.1 < z < 0.3$ and have a richness measure, $N_{200m}$, within $10 < N_{200m} < 190$.  The richness is defined as the number of red-sequence galaxies with $L > 0.4 L_*$ (in the $i$ band) within a projected radius of $R_{200m}$.  The catalog provides the BCG position (RA and DEC), photometric redshift, richness, BCG luminosity, and total luminosity of each cluster.  Applying the cluster detection method to mock catalogs suggests that the catalog should be $90\%$ pure and $85\%$ complete.  

Mass estimates of the clusters in the MaxBCG sample were derived by \citet{Sheldon2009} and \citet{Mandelbaum2008a} using weak gravitational lensing.  \citet{Johnston2007a} and \citet{Rozo2009} used those mass determinations to construct richness-mass ($N_{200m} - M_{500c}$) relations.  \citet{Rozo2009}, in particular, used the masses derived from \citet{Mandelbaum2008a} due in part to the authors' careful treatment of photometric redshift uncertainties \citep{Mandelbaum2008b}.  \citet{Rozo2009} also stacked the MaxBCG cluster catalog on X-ray maps from the ROSAT All-Sky Survey \citep{Voges1999} and used the $L_{\rm X} - M$ relation from \citet{Vikhlinin2009b} as a prior to inform their richness-mass relation.  Thus the richness-mass relation of \citet{Rozo2009} is expected to be consistent with an $L_{\rm X} - M$ relation from X-ray clusters.   \citet{Rozo2010} applied this richness-mass relation to the MaxBCG sample of optically-selected clusters and found a cosmological constraint on $\sigma_8$ of $\sigma_8(\Omega_m/0.25)^{0.41} = 0.832 \pm 0.033$ assuming a flat $\Lambda$CDM cosmology.  Throughout this work we define the $N_{200m} - M_{500c}$ relation as given by Eqs.~4, A20, and A21 of \citet{Rozo2009}.  

\subsection{The MCXC X-ray Cluster Catalog}\label{xray}

The Meta-Catalog of X-ray detected Clusters of galaxies (MCXC) is presented in \citet{Piffaretti2011}.  The MCXC cluster catalog is based on publicly available data from a number of different X-ray catalogs including the ROSAT All-Sky Survey and comprises 1743 clusters.  The catalog provides the position (RA and DEC), redshift, X-ray $0.1 - 2.4$ keV band luminosity ($L_{500c}$), mass ($M_{500c}$), and radius ($R_{500c}$) of each system.  The redshift distribution of this catalog goes from about 0.05 to 1.  The $L_{\rm X} - M$ relation derived from the MCXC clusters in \citet{Piffaretti2011} is consistent with that of \citet{Pratt2009} and \citet{Vikhlinin2009b}.  The \citet{Vikhlinin2009b} $L_{\rm X} - M$ relation was used to derive a cosmology constraint on $\sigma_8$ from a sample of X-ray clusters that is a subsample of the MCXC catalog.  From this analysis they found $\sigma_8(\Omega_m/0.25)^{0.47} = 0.813 \pm 0.013$ (stat) $\pm  0.024$ (sys) \citep{Vikhlinin2009b}.  Other authors have found similar constraints on $\sigma_8$ from ROSAT and other X-ray cluster samples \citep{Henry2009, Mantz2010}.

\subsection{Millimeter-wave Data from the Planck Satellite}\label{Planck}

Given that the optically-selected MaxBCG cluster catalog and the X-ray-selected MCXC cluster catalog yield consistent $L_{\rm X} - M$ relations and constraints on $\sigma_8$, one would expect both cluster samples to yield consistent $Y_{500c} - M_{500c}$ relations.\footnote{$Y_{500c}$ is the SZ flux within $R_{500c}$, dividing out the frequency dependence of the SZ signal.}  However, in a set of papers presented by the Planck collaboration \citep{Planck2011a,Planck2011b} it was found that the $Y_{500c} - M_{500c}$ relation for X-ray-selected clusters from the MCXC sample agrees with expectations, whereas the normalization of the $Y_{500c} - M_{500c}$ relation for the optically-selected MaxBCG sample was lower than expectations by about a factor of two.  For both cases, expectations were based on X-ray derived cluster profiles from \citet{Arnaud2010}.  For the latter case, the expectation also folded in the $N_{200m} - M_{500c}$ relation of \citet{Rozo2009}.  The {\it{Planck}} data used in the above analysis consists of six HFI channel millimeter-wave temperature maps as described in \citet{PlanckHFI2011a}.  This data set comprises the first ten months of the survey and covers the full sky.    

\subsection{Millimeter-wave Data from the Atacama Cosmology Telescope}\label{act}

The Atacama Cosmology Telescope is a six-meter telescope operating at an altitude of 5200 meters in the Atacama Desert of Chile.  The telescope site allows ACT to observe in both the northern and southern hemispheres.  In this work, we use millimeter-wave maps covering two regions of sky: one spanning 510 deg$^2$ over the celestial equator and one spanning 455 deg$^2$ in the southern hemisphere.  The Equatorial region consists of  a $4.5^\circ$-wide strip centered at a declination of $0^\circ$ and running from $20^{h}20^{m}$ through $0^{h}$ to $03^{h}50^{m}$.  The Southern region consists of a $7^\circ$ wide strip centered on $-53^\circ$ and extending from $00^{h}12^{m}$ to $7^{h}10^{m}$.  Both sky regions were observed over the 2008, 2009, and 2010 observing seasons at 148 and 218 GHz.  The Equatorial region overlaps the Sloan Digital Sky Survey (SDSS) Stripe 82 and thus overlaps 492 clusters in the MaxBCG catalog. The Equatorial plus Southern regions combined overlap 74 clusters in the MCXC catalog.\footnote{Only 6 clusters are in common between the 492 cluster MaxBCG sample and the 74 cluster MCXC sample.}   For a more detailed description of the ACT instrument, observations, and data reduction see \citet{Fowler2007, Swetz2011, Marriage2011, Das2011, Hajian2011, Dunner2012}.

\section{MEASUREMENTS OF SZ FLUX}\label{sec:szMeas}

\subsection{Multi-frequency Matched Filter}\label{sec:filter}

We use a multi-frequency matched filter to extract the thermal SZ signal from clusters as described in \citet{Haehnelt1996} and \citet{Melin2006}.  The filter in Fourier space is given by 

 \begin{equation}
{\boldsymbol{\Psi}}({\mathbf{k}}) =  \sigma^2_{\theta} ~[ {\mathbf{P}}({\mathbf{k}}) ]^{-1} \cdot {\boldsymbol{\tau}} ({\mathbf{k}}),
 \label{eq:filter}
 \end{equation}
 where ${\boldsymbol{\tau}}({\mathbf{k}})$ has the components
 
 \begin{equation}
 {\tau_{\nu}} ({\mathbf{k}}) =  \tau' ({\mathbf{k}}) ~{{j_{\nu}~B_\nu}} ({\mathbf{k}}).
 \end{equation}
Here ${{j_{\nu}}}$ is the frequency dependence of the thermal SZ signal for frequency $\nu$,  $\tau' ({\mathbf{k}})$ is the profile of the cluster in Fourier space, and ${{B_\nu}} ({\mathbf{k}})$ is the profile of the instrument beam in Fourier space.  ${\mathbf{P}}({\mathbf{k}})$ is the power spectrum of the noise, both astrophysical and instrumental.  The astrophysical noise sources for cluster detection include the primary lensed microwave background, radio galaxies, infrared galaxies, Galactic emission, and the SZ background from unresolved clusters, groups, and the intergalactic medium.  Since the power from the cluster thermal SZ signal is subdominant to these astrophysical sources (as evidenced by \citet{Lueker2010,Hall2010,Fowler2010,Das2011}), we approximate the power spectrum of the total noise as the power spectrum of the data itself. Here

 \begin{equation}
 \sigma^2_{\theta} = \Big[  \frac{1}{(2\pi)^2} \int d^2 k ~[  {\boldsymbol{\tau}} ({\mathbf{k}})  ]^t \cdot [ {\mathbf{P}}({\mathbf{k}}) ]^{-1} \cdot  [  {\boldsymbol{\tau}} ({\mathbf{k}})]  \Big]^{-1}
 \end{equation}
is the normalization of the filter that ensures an unbiased estimate of the cluster signal.

\subsection{SZ Model Template}\label{sec:szProfile}

We use for the filter's spatial template the empirical universal pressure profile of  \citet{Arnaud2010} derived from X-ray observations of the REXCESS cluster sample \citep{Bohringer2007}.  The three-dimensional pressure profile is given by

\begin{equation}
P^{3D}(r) \propto \frac{1}{x^{\gamma} (1+x^{\alpha})^{(\beta - \gamma)/\alpha}} \hspace{0.1cm},
\end{equation}
where $x=r/r_s$, $r_s = R_{500c}/c_{500}$, $c_{500} = 1.156$, $\alpha = 1.0620$, $\beta = 5.4807$, and $\gamma = 0.3292$.  The normalization of this profile is arbitrary for the purposes of the matched filter.  We essentially measure this normalization for each cluster when we apply this filter to our maps.  The SZ signal is given by the projected gas pressure, so we describe the filter template by integrating the three-dimensional profile above along the line of sight.  Thus

\begin{equation}
P^{2D}(\theta) = \int^{l_{max}}_{0} 2~P^{3D}\big(\sqrt{l^2 + \theta^2 D_A(z)^2}\big) dl,
\label{eq:profile}
\end{equation}
where $l_{max} = 5R_{500c} $ and $D_A(z)$ is the angular diameter distance.  The filter is truncated at $5R_{500c}/D_A(z) = 5\theta_{500c}$, which contains over $95\%$ of the signal, as was done in \citet{Melin2011,Planck2011a,Planck2011b}.

\subsection{Application of Filter}\label{sec:application}

Before filtering the maps, we establish uniform noise properties by creating an effective weight map that has pixel-wise effective weights given by $w_{\rm{eff}} = (\frac{1}{w_1} + \frac{1}{w_2})^{-1}$ where $w_i$ is the pixel weight for the $i$th frequency.  Here weight is defined as the number of observations per pixel normalized by the observations per pixel in the deepest part of the map.  We multiply the 148 and 218 GHz ACT data maps pixel-wise by the square root of the effective weight map.  After we apply the filter to create a filtered map, we divide the filtered map pixel-wise by the square root of the effective weight map.

\begin{figure}[t!]
\begin{center}
\epsscale{1.0}
\plotone{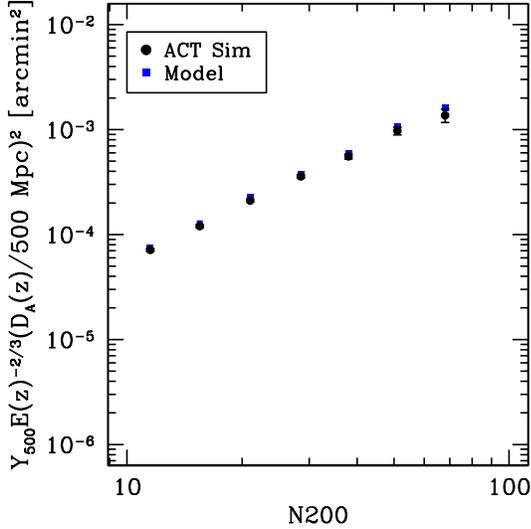}
\caption{Recovered $Y_{500}$ values from 447 simulated clusters embedded in ACT 148 and 218 GHz maps (black circles).  These clusters were simulated to match the properties of the clusters in the MaxBCG catalog \citep{Koester2007b} that overlap the ACT equatorial region.  The simulated clusters were placed at random locations within the ACT maps.   The input $Y_{500}$ values are shown as blue squares.}
\label{fig:sim1}
\end{center}
\end{figure}

\begin{figure}[b!]
\epsscale{1.0}
\plotone{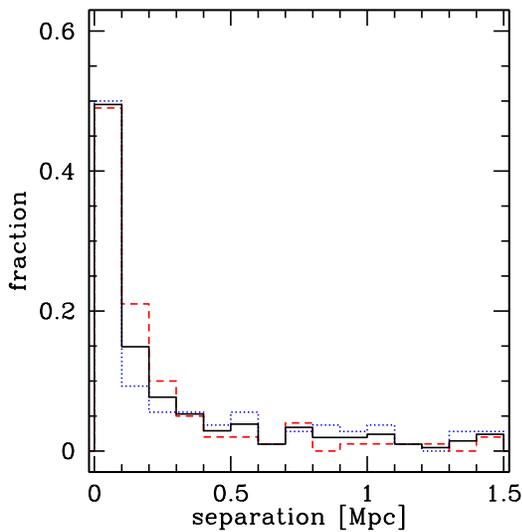}
\caption{The black solid line shows the distribution of offsets between BCG and X-ray gas peak from 208 clusters found in both the MaxBCG and MCXC cluster catalogs.  The red dashed line shows the distribution for the subsample of rich clusters ($N_{200m} \geq 35$).  The blue dotted line shows the same for the subsample of poor clusters ($N_{200m} < 35$).  There are 100 and 108 clusters in each subsample respectively.}
\label{fig:distribution}
\end{figure}

\begin{figure}[t!]
\begin{center}
\epsscale{1.0}
\plotone{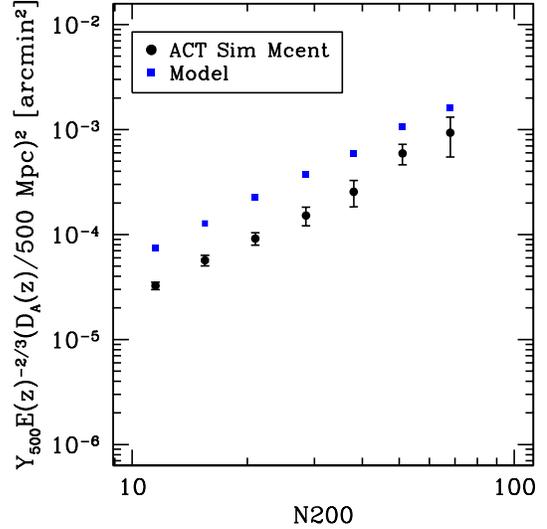}
\caption{Black circles show recovered $Y_{500}$ values from 446 embedded simulated clusters where the assumed cluster centers used for $Y_{500}$ recovery are offset from the input centers with a random distribution given by the black solid line in Figure \ref{fig:distribution}.  "Mcent" stands for miscentered.  The input values are shown as blue squares.}
\label{fig:sim2}
\end{center}
\end{figure}

We apply the matched filter above following the procedure given in  \citet{Planck2011b}.  For each cluster in the MaxBCG catalog that falls in the ACT coverage region, we create a unique matched filter using the $N_{200m} - M_{500c}$ relation of \citet{Rozo2009} to determine each cluster's $M_{500c}$ and subsequently $R_{500c}$ from the $N_{200m}$ value given in the catalog.  We also derive $D_{A}(z)$ for each cluster from its photometric redshift.  When the filter is applied to the map, the pixel coincident with the location of the cluster center in the filtered map should have a value equal to the normalization of the two-dimensional SZ template given by Eq.~\ref{eq:profile}.  To simplify extraction of the desired quantity, we normalize $P^{2D}(\theta)$ itself to equal unity when integrated over $\theta$ in two dimensions from zero to $5\theta_{500c}$.  Thus the pixel value recovered at the cluster center position after applying the filter is $Y^{cyl}_{5\theta_{500c}}$.  Here $Y^{cyl}_{5\theta_{500c}} D_{A}(z)^2 = Y^{cyl}_{5R_{500c}}$,  where $Y^{cyl}_{5R_{500c}}$ is the integrated projected SZ signal within a cylinder of radius $5R_{500c}$.  We use a geometric factor of $Y^{sph}_{R_{500c}} = (0.986/1.814)Y^{cyl}_{5R_{500c}}$ given in Appendix A of \citet{Melin2011} to convert from $Y^{cyl}_{5R_{500c}}$ to $Y^{sph}_{R_{500c}}$, the integrated SZ flux within a sphere of radius $R_{500c}$.  Throughout this work we plot ${\tilde{Y}_{500}} \equiv Y_{500} E^{-2/3}(z) (D_{A}(z)/500~{\rm{Mpc}})^2$, where $Y_{500} = Y^{sph}_{\theta_{500c}}$, as in  \citet{Planck2011b}.

\section{SIMULATED ACT SZ SIGNALS}\label{sec:sims}

\subsection{Embedding Simulated Clusters in ACT Maps}\label{sec:embed}

In order to test the analysis pipeline which applies the SZ extraction procedure discussed above, we use simulated clusters embedded within the ACT data maps at random locations.  Using the information in the MaxBCG catalog for the 492 clusters that fall within the ACT Equatorial region, we create a unique SZ profile for each cluster using Eq.~\ref{eq:profile} above and the cluster $R_{500c}$ and $z$ given in the catalog.  We then add each simulated SZ cluster to the 148 and 218 GHz ACT Equatorial maps, placing it at a random location and scaling the thermal SZ signal to give it the appropriate frequency dependence in each map.  The simulated SZ signal is also convolved with the appropriate ACT beam prior to embedding it within each ACT map.  We then exclude any simulated clusters from further analysis that happen to be within 5\arcmin~of a point source detected at greater than 5$\sigma$ in either the 148 or 218 GHz maps.  We also exclude any clusters that are within 10\arcmin~from the edge of the map.  As an additional cut, we exclude all clusters that are in noisy parts of the map where the local value of the effective weight map is less than $15\%$ of the maximum value. These cuts leave 447 clusters.  Similarly below, when we extract the SZ signal from the real cluster positions using the MaxBCG and MCXC catalogs, we apply the same cuts discussed above.

In Figure \ref{fig:sim1}, we show the results of this SZ extraction procedure.  Here the simulated clusters are binned by the optical richness given in the MaxBCG catalog.  The blue squares show the input model, and the black circles show the recovered signal of the simulated clusters embedded at random locations within the ACT maps.  The error bars are given by $\sigma/\sqrt{N}$ where $\sigma^2$ is the variance of ${\tilde{Y}_{500}}$ in each richness bin, and $N$ is the number of clusters in each richness bin.  The variance dominates the uncertainty in each bin, as demonstrated in Figure 4 of  \citet{Planck2011b}.\footnote{Since no scatter in the $N_{200m} - M_{500c}$ relation has been included here, these error bars reflect only the SZ flux recovery error.}  There is good agreement between input and recovered signals.

\begin{figure}[t!]
\epsscale{1.0}
\plotone{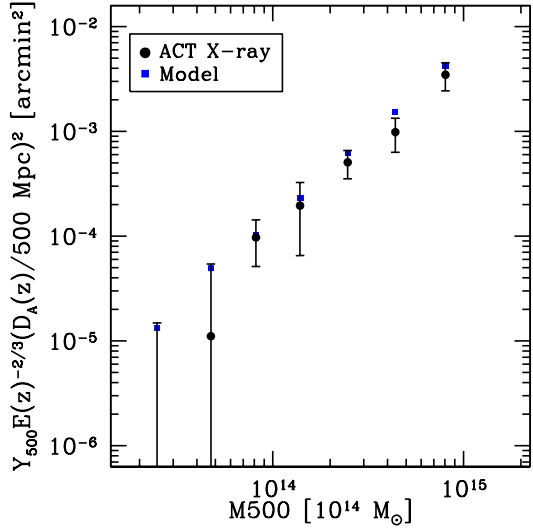}
\caption{Measured $Y_{500}$ values for 52 MCXC X-ray-selected clusters \citep{Piffaretti2011} that fall within the ACT equatorial and southern survey regions (black circles).  Also shown are expected $Y_{500}$ values based on measured cluster X-ray properties (blue squares).  A cluster profile model from \citet{Arnaud2010} was assumed for determining both measured and expected $Y_{500}$ values. \label{fig:X-ray}}
\end{figure}

\subsection{Effect of Cluster Miscentering}\label{sec:miscenter}

The exercise above shows we should expect excellent recovery of the SZ flux for clusters given in the MaxBCG catalog provided the $N_{200m} - M_{500c}$ and $M_{500c} - Y_{500c}$ relations are correct, and the cluster properties (position, redshift, $N_{200m}$) listed in the catalog are accurate.  However, one source of uncertainty identified in \citet{Johnston2007b} (section 4.3) is the positional accuracy of the cluster center.  In the MaxBCG catalog, the cluster position is given by the location of the Brightest Cluster Galaxy (BCG).   \citet{Johnston2007b} suggested two reasons why the BCG found by the MaxBCG cluster identification algorithm may be offset from the true cluster center.  One is that the true BCG may be intrinsically offset from the dark matter center or center of the gas concentration presumably due to unrelaxed behavior (e.g., mergers).  Another reason may be that the BCG could be misidentified by the cluster finder.   \citet{Johnston2007b} explore this latter effect with mock optical cluster catalogs and find that a richness-dependent fraction of the clusters have accurately identified BCGs while the rest are miscentered due to BCG misidentification following a Rayleigh distribution with a scale parameter $\sigma$ equal to $\sigma_{\rm{misc}} = 0.42 h^{-1} {\rm{Mpc}}$.  The distribution of the intrinsic BCG offset from the gas center is unknown.\footnote{Note that this offset does not arise from pointing uncertainties in optical, X-ray, or millimeter-wave instruments.  It is due to either BCG misidentification or cluster astrophysics.}

We explore the potential offset between BCG and gas center by studying the subset of clusters in common to both the MaxBCG and MCXC catalogs.  From the full catalogs (which have 13,823 and 1743 clusters respectively), we identify 208 clusters that are in common in both catalogs.  Here we define a cluster as matched in both catalogs when the identified cluster redshifts are within $\Delta z < 0.015$ and the projected physical separation of the identified cluster centers is less than 1.5 Mpc.  Using these 208 clusters, we plot the fraction of clusters as a function of separation between BCG and X-ray peak in Figure~\ref{fig:distribution}.  The solid black line shows the offset distribution for the 208 clusters.  Half of the clusters have offsets less than 0.1 Mpc, while the other half have a roughly flat offset distribution between 0.1 and 1.5 Mpc.\footnote{We note that the choice of 1.5 Mpc is somewhat arbitrary.  When we allow matches within 1 Mpc, we find 189 clusters.    We also find that if we allow matches within 3 Mpc, Figure~\ref{fig:distribution} plateaus instead of dropping to zero at large separation.  The plateau is due to poor clusters, as the distribution of rich ones does tend to zero.  The size of a cluster ($R_{200m}$) is usually less than 2 Mpc, so this suggests that some of these poor cluster matches may be spurious.}

\begin{figure}[t!]
\epsscale{1.0}
\plotone{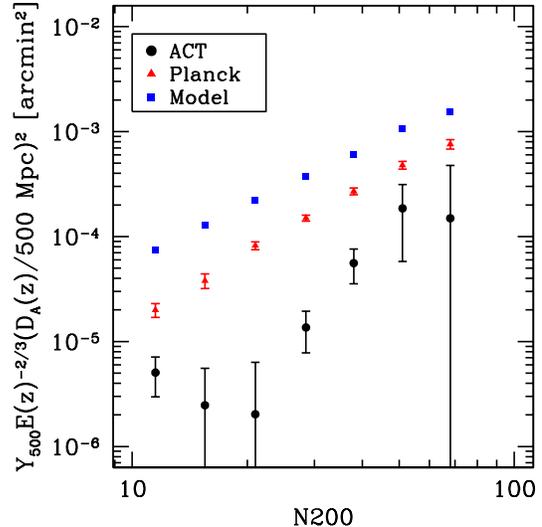}
\caption{Measured $Y_{500}$ values for 474 MaxBCG optically-selected clusters that fall within the ACT equatorial survey region (black circles).  Expected $Y_{500}$ values are shown as blue squares.  Both measured and expected values assume the $N_{200m} - M_{500c}$ relation from \citet{Rozo2009} and the  \citet{Arnaud2010} cluster profile.  Red triangles are the measured values from the {\em{{\it{Planck}}}} satellite for a sample of 13,104 MaxBCG clusters  \citep{Planck2011b}. \label{fig:Optical}}
\end{figure}

The 208 clusters were then divided into rich clusters (with $N_{200m} \geq 35$) and poor clusters (with $N_{200m} < 35$).  A richness cut of $N_{200m} = 35$ divides the 208 clusters into roughly even subsamples of $\sim 100$ clusters each.  The dashed red line in Figure \ref{fig:distribution} shows the offset distribution for the rich clusters, and the dotted blue line shows the distribution for the poor clusters.  Similar distributions are found for both subsamples, with the poor clusters showing slightly more offset.

Using the simulations described above, we explore the effect of cluster miscentering on SZ flux recovery.  We use the same simulated clusters embedded at random positions in the 148 and 218 GHz ACT maps as before.  However, when recovering the SZ fluxes, we use positions for the clusters that differ from the true cluster gas centers with a random distribution given by the black solid line in Figure \ref{fig:distribution}.  We also allow for $10\%$ of the clusters to be false detections to match the purity of the MaxBCG catalog \citep{Koester2007a}.  Figure \ref{fig:sim2} shows the result of the SZ recovery process.  The black circles show the recovered $Y_{500c}$ values for simulated clusters embedded in ACT data.  To give a sense of the relation between Mpc and arcminutes, for the redshift range of the MaxBCG cluster sample ($z\in(0.1, 0.3)$), 0.5 Mpc corresponds to about 2\arcmin~to 4.5\arcmin.  Given the ACT beam size (1.4\arcmin~at 148 GHz and 1\arcmin~at 218 GHz), we expect a decrease in the recovered $Y_{500c}$ signal due to the amount of miscentering shown in Figure \ref{fig:sim2}.  How much the recovered SZ signal decreases depends on the noise properties of the ACT data, which differ significantly from pure white noise.   We discuss this further in Section \ref{sec:discuss}.

\begin{figure}[t!]
\epsscale{1.0}
\plotone{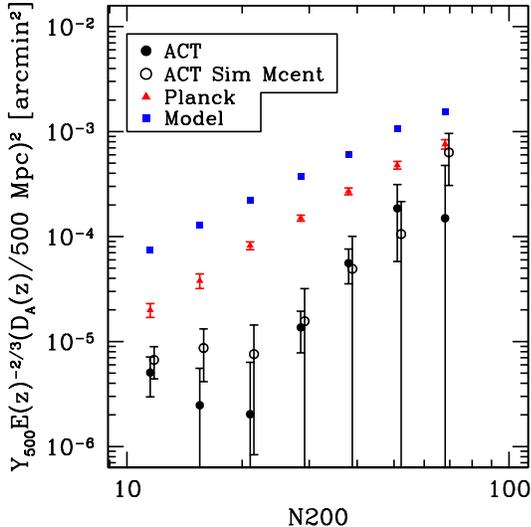}
\caption{The black open circles show the measured $Y_{500}$ values from simulations with an offset between identified cluster center and gas concentration center given by a uniform random distribution between 0 and 1.5 Mpc.  The solid black circles, blue squares, and red triangles are the same as in Figure \ref{fig:Optical}. The values from simulations have been shifted in the x-axis for clarity.
\label{fig:comp}}
\end{figure}

\section{MEASURED ACT SZ SIGNALS}\label{sec:data}

\subsection{Stacking X-ray Selected Clusters}\label{sec:xray}

Within the 510 deg$^2$ ACT Equatorial region and the 455 deg$^2$ ACT Southern region are located 74 clusters found in the MCXC catalog of X-ray-selected clusters  \citep{Piffaretti2011}.  After making the cuts discussed in Section \ref{sec:embed} to exclude clusters near bright point sources, near map edges, or in very noisy parts of the map, 52 MCXC clusters remain.  Using the $R_{500c}$, $M_{500c}$, and redshift of each cluster and the projected SZ profile given in Eq.~\ref{eq:profile}, we calculate the expected mean $Y_{500c}$ values in each $M_{500c}$ bin shown as the blue squares in Figure \ref{fig:X-ray}.  The black circles in Figure \ref{fig:X-ray} show the mean of the recovered $Y_{500c}$ values using the multi-frequency matched filter given in Eq.~\ref{eq:filter} and the projected SZ profile created uniquely for each cluster.  The error bars are the error on the mean given by $\sigma/\sqrt{N}$, where $\sigma^2$ is the variance and $N$ is the number of clusters in each bin.  Figure \ref{fig:X-ray} shows overall agreement between expected and recovered SZ signals for the X-ray-selected clusters. The reduced chi-squared is 0.76 using 7 degrees of freedom.  This is consistent with the agreement found by \citet{Planck2011a} for a larger sample of X-ray-selected clusters.

\subsection{Stacking Optically Selected Clusters}\label{sec:optical}

\begin{figure*}[t!]
\begin{center}
\includegraphics[scale=0.4]{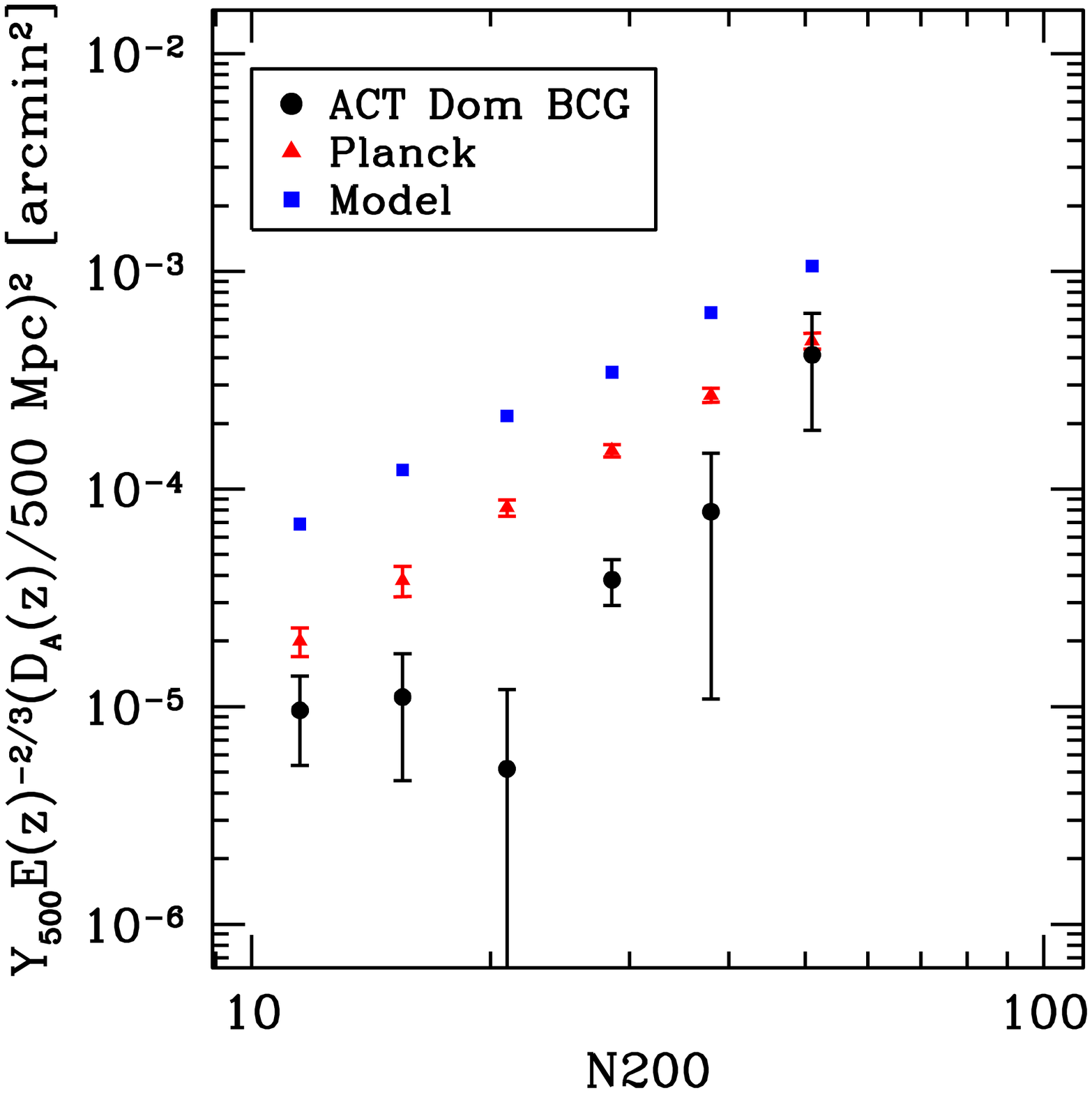}
\hspace{0.1cm}
\includegraphics[scale=0.4]{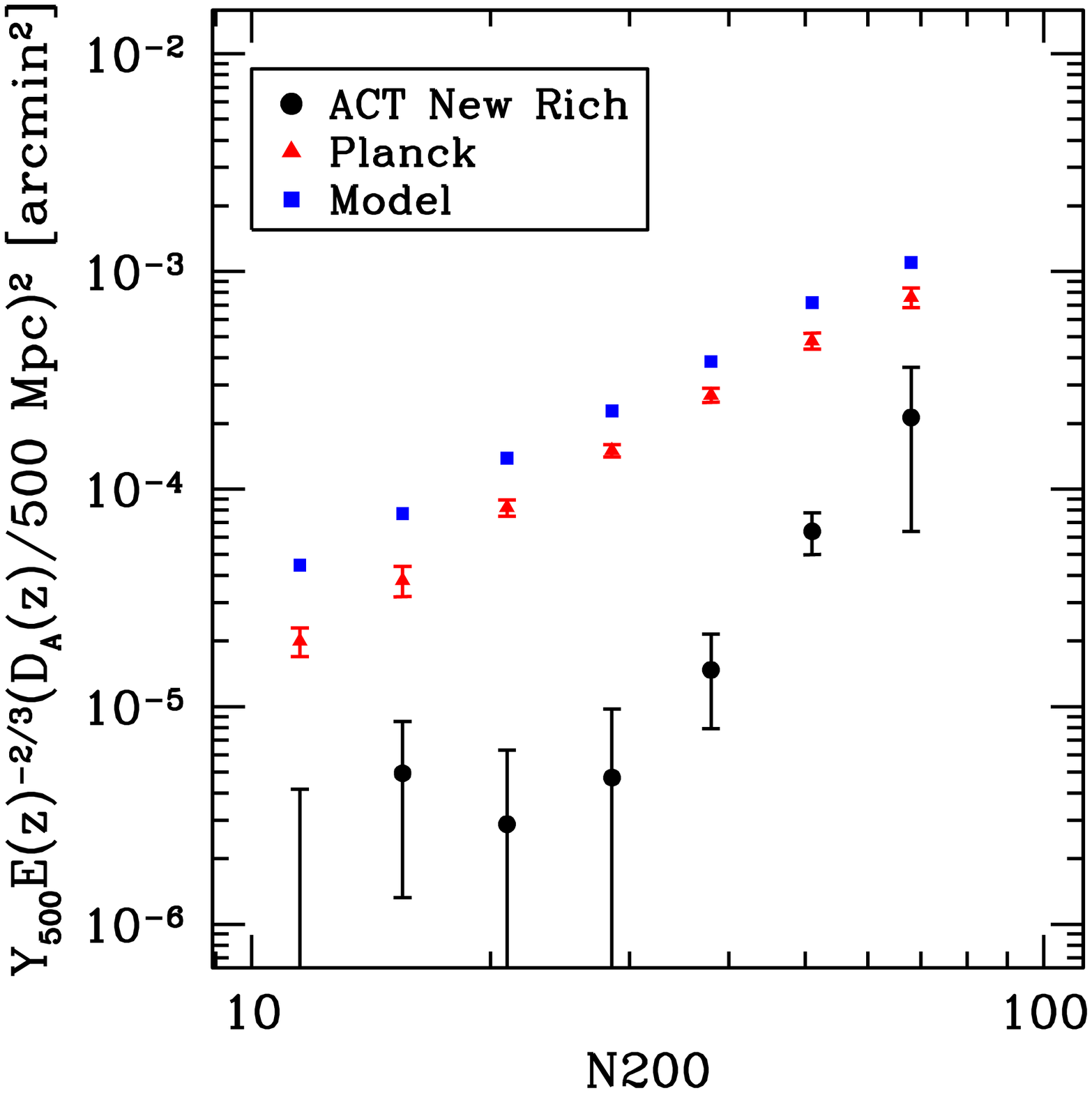}
\caption{ {\emph{Left (a)}: Shown are the recovered $Y_{500}$ values as in Figure \ref{fig:Optical}, except using a subsample of 126 "BCG dominant" clusters from the 474 MaxBCG clusters used above.   Both the measured values (black circles) and expected values (blue squares) change using the new subsample as compared to Figure \ref{fig:Optical}. The red triangles are the measured values from {\em{{\it{Planck}}}} for the 13,104 MaxBCG sample.}  {\emph{Right (b):} Shown are the same points as in Figure \ref{fig:Optical}, however, using the new richness measure for the MaxBCG cluster sample given in \citet{Rykoff2011}.   Note that both the measured values (black circles) and expected values (blue squares) change using the new richness measure as compared to Figure \ref{fig:Optical}.  The red triangles are the {\em{{\it{Planck}}}} measured values using the {\it{old}} richness measure for reference.} }
\label{fig:BcgDomNewRich}
\end{center}
\end{figure*}

In the ACT Equatorial region there are 492 MaxBCG clusters.  This reduces to 474 clusters once the above mentioned cuts are made.\footnote{Note that the ACT region is centered on Stripe 82, and extends beyond it. So fewer real clusters are cut than simulated clusters because the real clusters are only in Stripe 82, and thus are located towards the central part of the ACT map which has lower noise.}  Using the $N_{200m} - M_{500c}$ relation of \citet{Rozo2009} and the $M_{500c} - Y_{500c}$ relation of \citet{Arnaud2010}, we find the expected and recovered $Y_{500c}$ values of the MaxBCG clusters using the method described in Section \ref{sec:szMeas}.  Figure \ref{fig:Optical} shows the expected values as blue squares and the recovered values as black circles.  The recovered signal is significantly lower than the expected signal as well as the signal recovered by {\it{Planck}} from 13,104 MaxBCG clusters (red triangles) \citep{Planck2011b}.  

From Figure \ref{fig:sim2}, we see that some amount of offset between BCG and gas peak can result in a lower measured signal than expected.  However, this figure also shows that the offset distribution given by Figure \ref{fig:distribution}  does not result in a low enough measured signal to explain the measurement shown in Figure \ref{fig:Optical}.  To investigate the amount of offset necessary to match the ACT measured $Y_{500}$ values shown in Figure \ref{fig:Optical}, we redo the analysis shown in Figure \ref{fig:sim2}.  However, this time we use an offset distribution that is uniformly random between 0 and 1.5 Mpc.  We also again allow for $10\%$ false detections. The results are shown as the open black circles in Figure \ref{fig:comp}.  The solid black circles, blue squares, and red triangles in Figure \ref{fig:comp} are the same as in Figure \ref{fig:Optical}.  We see that this amount of offset between BCG and gas peak roughly matches the measured values.  An extensive scan of offset distributions is beyond the scope of this work, but it may be that a more complex or refined distribution could give a better fit to the measurements. 

\subsection{Stacking Optical Clusters Using BCG Dominant Subsample and New Richness Measure}\label{sec:optical2}

Using a subsample of clusters with "dominant BCGs" we examine whether the expected SZ signal is closer to the measured values.  Such a subsample may more closely correspond to an X-ray-selected subsample, and with such a subsample \citet{Planck2011b} found better agreement between model and measurement.  We follow the definition of "BCG dominant" used by  \citet{Planck2011b} which is defined relative to the quantity $L_{\rm{BCG}} /(L_{\rm{tot}} - L_{\rm{BCG}})$.  Here $L_{\rm{tot}}$ and $L_{\rm{BCG}}$ are the R-band luminosities of the cluster and cluster BCG respectively.  For a "BCG dominant" cluster, this ratio is larger than the average ratio for a given richness bin.  From the sample of 474 MaxBCG clusters used above, 126 are "BCG dominant".  Figure \ref{fig:BcgDomNewRich}a shows the recovered $Y_{500}$ values versus the model expectation for this subsample.  The {\it{Planck}} measurements of the sample of 13,104 MaxBCG clusters (not a subsample) is included for reference.

Recently a new measure of cluster richness was developed with less scatter than the measure presented in \citet{Koester2007b} \citep{Rykoff2011}.  We test whether using this new richness measure will yield differing results with regard to measured versus expected SZ signal.  A catalog with a new richness measure assigned to each MaxBCG cluster is available online.\footnote{http://kipac.stanford.edu/maxbcg}  While for the previous richness measure the cluster richness-mass relation was calibrated  using weak lensing, this new relation was calibrated with an abundance matching technique \citep{Rykoff2011}.  Calibration of this new measure via weak lensing is still in progress.  We use the richness-mass relation given in Eq.~B6 of  \citet{Rykoff2011} to determine $M_{500c}$ and subsequently $R_{500c}$.  Figure \ref{fig:BcgDomNewRich}b shows the measured and expected $Y_{500}$ values using this new richness measure for the 474 MaxBCG clusters.  The red triangles show the {\it{Planck}} measurements using the {\it{old}} richness measure for reference.  Note that {\it{Planck}} would have different results if they use the new richness measure.
 
\subsection{Contamination from Infrared and Radio Galaxies}\label{sec:contam}

\begin{figure}[b!]
\epsscale{1.0}
\plotone{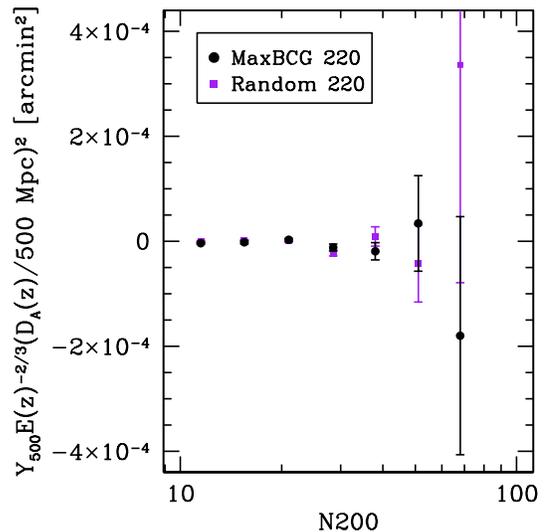}
\caption{Measured $Y_{500}$ values from 474 MaxBCG clusters using the 218 GHz ACT equatorial map alone (black circles).  For comparison is shown the same $Y_{500}$ recovery procedure performed at 474 random locations in the 218 GHz map (purple squares).}
\label{fig:contam}
\end{figure}

To investigate whether infrared galaxies may be reducing the SZ decrement at the MaxBCG cluster positions, we recover the $Y_{500}$ values from the 474 MaxBCG clusters studied above using the single band 218 GHz ACT map alone.  We compare that to $Y_{500}$ values extracted at 474 random positions within the 218 GHz map.\footnote{For these measurements of $Y_{500}$, we do not divide out the amplitude of the frequency dependence of the SZ signal, which is close to zero at the null frequency.  So these are really measurements of $-\Delta T/T_{\rm{cmb}}$.} A positive correlation between MaxBCG clusters and infrared galaxies would result in negative $Y_{500}$ values compared to the random sample.  We find that both the MaxBCG cluster sample and the random sample have an SZ flux consistent with zero in the 218 GHz map, and we do not detect any significant excess of infrared signal correlated with the MaxBCG sample.  This is shown in Figure~\ref{fig:contam}.

We also cross-correlate the MaxBCG cluster catalog with the VLA FIRST catalog of radio sources to investigate how much radio galaxies may be reducing the measured SZ decrement.\footnote{http://sundog.stsci.edu/first/catalogs/readme\_12feb16.html}  The FIRST survey uses the NRAO Very Large Array and covers over 10,000 square degrees of sky to a sensitivity of about 1 mJy at 1.4 GHz.  This survey also overlaps with the Sloan Digital Sky Survey.  We cross-correlate to find the fraction of 474 MaxBCG clusters that have a radio source above a given flux threshold, within 5\arcmin~of the identified cluster position.  Assuming a typical spectral index for radio sources of -0.7 \citep[e.g.,][]{Condon1984,Lin2009}, we choose a flux threshold cut of 50 mJy at 1.4 GHz to yield sources above 2 mJy at 150 GHz.  Such a source convolved with the ACT beam would have a temperature increment of about $30 \mu K$ at 148 GHz.  Since the typical SZ signal from a cluster is about $100 \mu K$ to within a factor of a few when smoothed with the ACT beam, a 2 mJy radio source would start to significantly reduce the SZ decrement.  We find that about $10\%$ of the 474 MaxBCG clusters investigated above have such a radio source within 5\arcmin~of its identified center.  This small correlation is not enough to explain the large discrepancy between measured and expected SZ signals shown in Figure \ref{fig:Optical}, although it may be a contributing factor.

\begin{figure}[t!]
\epsscale{1.0}
\plotone{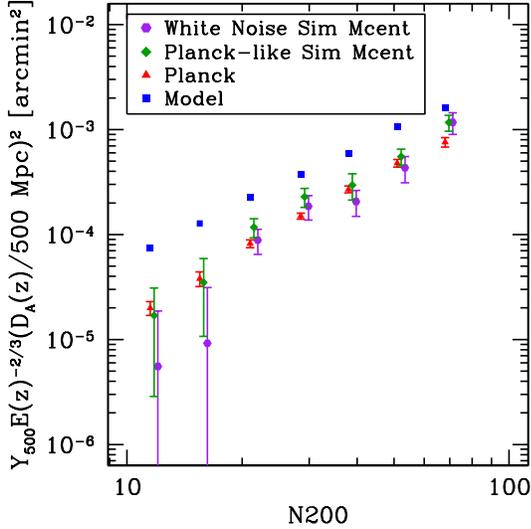}
\caption{Recovered $Y_{500}$ values from 984 simulated clusters embedded in simulated CMB maps at 148 and 218 GHz with {\em{{\it{Planck}}}}-like noise (green diamonds) or the corresponding level of white-noise (purple hexagons).  The maps were convolved with ACT beams as well as a  5\arcmin~Gaussian beam, so that they correspond to {\em{{\it{Planck}}}} resolution.  The recovered positions are offset from the true positions with a uniform random distribution between 0 and 1.5 Mpc.  The input $Y_{500}$ values are shown as blue squares, and red triangles are the measured values from the {\em{{\it{Planck}}}} satellite for a sample of 13,104 MaxBCG clusters  \citep{Planck2011b}.\label{fig:discuss}}
\end{figure}

 \section{SIMULATED PLANCK SZ SIGNALS}\label{sec:planck}
 
To investigate what {\em{{\it{Planck}}}}'s measured SZ signal would be if there existed the amount of offset modeled in Figure~\ref{fig:comp}, we again use simulations.  We embed 492 simulated MaxBCG clusters in two sets of simulated CMB maps at 148 and 218 GHz that have {\em{{\it{Planck}}}}-like instrument noise added to them.\footnote{We doubled the number of maps and thus the cluster sample to shrink the error bars.  This results in 984 embedded clusters in total.}  We model the {\em{{\it{Planck}}}} noise using the the noise power spectra at 143 and 217 GHz shown in Figure 35 of \citet{PlanckHFI2011a}.  We allow for an offset between the cluster SZ peaks and the identified cluster centers that has a uniform random distribution between 0 and 1.5 Mpc, analogous to Figure~\ref{fig:comp}.  We also convolve both maps with ACT beams and a 5\arcmin~Gaussian beam to approximate {\em{{\it{Planck}}}} resolution.  The results of the extracted $Y_{500}$ values are shown as green diamonds in Figure~\ref{fig:discuss}.  We also show the case where the simulated clusters are embedded in simulated CMB plus white noise maps, using white noise levels that are similar to {\em{{\it{Planck}}}} noise levels.  These results are shown as purple hexagons in Figure~\ref{fig:discuss}.

{\em{{\it{Planck}}}}-like noise is nearly white at these frequencies, so there is not much difference between the {\em{{\it{Planck}}}}-like case and the white noise case in Figure~\ref{fig:discuss}.  The results of the {\em{{\it{Planck}}}}-like case in Figure~\ref{fig:discuss} are also very different from those of the simulated ACT case shown by the open black circles in Figure~\ref{fig:comp}.  The two differences between the simulations are the beam and the noise.  In the ACT data there is 1/f noise, atmospheric noise, and noise from the primary microwave background.\footnote{The instrumental noise in the ACT 218 GHz map is not low enough to remove all the primary microwave background signal from the 148 GHz map.}  All of this results in a redder noise spectrum than {\em{{\it{Planck}}}}'s.  The presence of red noise in the maps causes the matched filter to suppress more power on large scales than would be the case for white noise. This causes the filter to return a lower signal than actually exists if the signal is extracted from a position that is offset from the cluster center.  If the signal is extracted at the cluster center, however, the matched filter will return the correct signal regardless of whether the noise is white or somewhat red (as can be inferred from Figure~\ref{fig:sim1}).\footnote{Note that estimating the noise power spectrum as CMB plus white noise when filtering the ACT data results in very large errorbars since the matched filter is no longer optimal. High-pass filtering the ACT maps enough to minimize the effect of the red noise causes ringing of the true cluster signal, which also results in a low measurement if the signal is extracted from a position offset from the center.}  

\begin{figure}[b!]
\epsscale{1.0}
\plotone{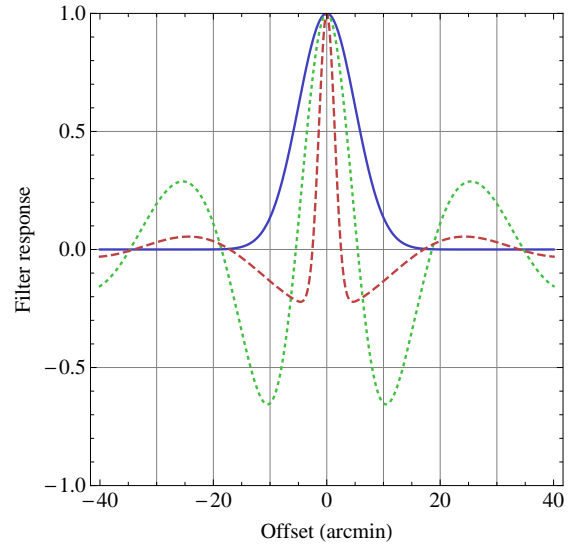}
\caption{Shown is the filter response for differing beams and noise models.  The blue solid curve is a 5\arcmin~Gaussian profile, which roughly approximates the filter response for a cluster profile convolved with the {\em{{\it{Planck}}}} beam since {\em{{\it{Planck}}}}'s noise is close to white (see Eq.~\ref{eq:filter}).  The red dashed curve shows a 1.4\arcmin~Gaussian profile, which has been Fourier transformed and has had all Fourier modes set to zero for $l<2000$, roughly analogous to the effect of the ACT noise.  The profile was then Fourier transformed back.  The green dotted curve shows the same as the red curve except for a 5\arcmin~Gaussian profile.  \label{fig:filter}}
\end{figure}

We demonstrate this effect in Figure~\ref{fig:filter}, where the blue solid curve is a 5\arcmin~Gaussian profile.  This is roughly the filter response for a cluster profile convolved with the {\em{{\it{Planck}}}} beam as {\em{{\it{Planck}}}}'s noise is close to white (see Eq.~\ref{eq:filter}).  The red dashed curve shows a 1.4\arcmin~Gaussian profile, which had been Fourier transformed and had all Fourier modes set to zero for $l<2000$, roughly analogous to the effect of the ACT noise.  The profile was then Fourier transformed back.  One can see how a low flux recovery results from an offset from the center.  The green dotted curve shows the same as the red curve except for a 5\arcmin~Gaussian profile.  The low signal from a miscentered position is still apparent.  This is why merely smoothing the ACT maps to match the {\em{{\it{Planck}}}} beam would not allow us to reproduce the {\em{{\it{Planck}}}} measurement.  

It is interesting to see from Figure~\ref{fig:discuss} that the amount of offset modeled in Figure~\ref{fig:comp} to match the ACT data, can also explain the discrepancy found by {\em{{\it{Planck}}}}.  If this amount of miscentering actually exists, then it would be the sole explanation of the discrepancies.  However, this amount of offset between BCGs and SZ peaks is much larger than the distribution shown in Figure~\ref{fig:distribution}.  So it may be that the SZ signal is intrinsically low by some amount or that the optical weak-lensing mass calibration is biased high.  It is also possible that the fraction of false detections in the optically-selected sample is larger than $10\%$, or some non-zero amount of radio or infrared galaxy contamination is filling in the SZ decrements (as discussed in Section~\ref{sec:contam}).  If so, these effects would serve to lower the measured SZ signal by the same amount for both ACT and {\em{{\it{Planck}}}}.   In that case, a smaller amount of miscentering would be required to match the remaining discrepancy found by these instruments.   It is thus likely that a combination of effects may be at work resulting in the measurements found by ACT and {\em{{\it{Planck}}}}.  

In this vein, we have checked with simulations whether an intrinsic SZ signal as low as measured by {{\em{{\it{Planck}}},} combined with the distribution of BCG/SZ peak separations shown in Figure~\ref{fig:distribution}, can roughly match the SZ signal measured by ACT.  We did this by lowering the SZ signal of the simulated clusters until they matched the {{\em{{\it{Planck}}}} measurements, and then recovered this signal using wrong gas peak positions given by the distribution in Figure~\ref{fig:distribution}, as done in Section~\ref{sec:miscenter}.  We find that the recovered simulated SZ signal is at least a factor of two higher than measured by ACT.
    
 \section{DISCUSSION}\label{sec:discuss}

From the analysis of simulated clusters embedded in ACT data presented in Figure~\ref{fig:sim1}, we expect robust recovery of the SZ flux from optically-selected MaxBCG clusters.  This is assuming that the position of the center of the gas concentration is known.  From the analysis of 52 MCXC clusters that fall within the ACT Equatorial and Southern survey regions shown in Figure~\ref{fig:X-ray}, we find agreement between the expected and measured $Y_{500c} - M_{500c}$ relation.  Such agreement is consistent with that found in \citet{Planck2011a}.  Both of these figures taken together give confidence in the SZ flux recovery pipeline and the analysis of the ACT maps.

In Figure~\ref{fig:Optical}, we find that the recovered ACT SZ flux from 474 optically-selected MaxBCG clusters is lower than both the model expectations and the measured {\em{{\it{Planck}}}} values given in \citet{Planck2011b}.  Since we expect from Figure~\ref{fig:distribution} some offset between the positions of the BCGs and the centers of the gas concentrations, we explore this possibility in more detail.   Given the difference in resolution and noise properties between ACT and {\em{{\it{Planck}}}}, such an offset would result in different measured SZ signals between the two instruments (see Figure~\ref{fig:filter}).

Figure~\ref{fig:distribution} gives the offset distribution for clusters in common between the full MaxBCG and MCXC cluster catalogs.  Using simulations, Figure~\ref{fig:sim2} shows that this amount of offset is not enough to explain the measured SZ signal shown in Figure~\ref{fig:Optical}.  Modeling the offset using the \citet{Johnston2007b} distribution, which only includes misidentification of the BCG by the optical cluster finding algorithm, produces a slightly smaller amount of discrepancy between expected and measured SZ signals than what is shown in Figure~\ref{fig:sim2}.  To match the measured SZ signal found by ACT requires an offset distribution that is significantly larger.  A distribution that is uniformly random between 0 and 1.5 Mpc gives a better fit to the ACT data, as shown in  Figure~\ref{fig:comp}.  This amount of offset also can explain the {\em{{\it{Planck}}}} measured discrepancy as shown in Figure~\ref{fig:discuss}.

It is possible that the subsample of clusters in common between MCXC and MaxBCG catalogs is not representative of the full MaxBCG cluster sample.  This subsample is in fact special in that these clusters are found using both optical and X-ray selection techniques.  When {\em{{\it{Planck}}}} measured the SZ signal for such a subsample, no significant discrepancy between the measured and expected SZ signals was found \citep{Planck2011b}.  This is in contrast to their results for the full optically-selected sample.  This suggests that the subsample of clusters in common to both catalogs is not representative of the full optically-selected sample, at least in some regard. 

We also assume throughout this work that the \citet{Arnaud2010} profile accurately describes the gas profiles of clusters in the MCXC and MaxBCG catalogs.  The good agreement we find in Figure~\ref{fig:X-ray} and also shown in \citet{Planck2011a} suggests that this profile works well for clusters in the MCXC sample.  However, this profile may not hold for the optically-selected MaxBCG sample.  In particular, to measure a lower SZ signal with ACT than found by {\em{{\it{Planck}}}} would require that the gas distribution is wider (more spread out) than that given by the  \citet{Arnaud2010} profile. In such a case, given the finer resolution of ACT, it is conceivable that ACT could measure a lower signal than {\em{{\it{Planck}}}} when the SZ signal is extracted using an incorrect narrower profile for both datasets.  However, since we expect clusters to be in hydrostatic equilibrium to first order, the consequence of a wider gas distribution is a lower total mass within $M_{500c}$ if the gas mass fraction is fixed. This would imply lower normalizations of the $L_{\rm X} - M_{500c}$ and $N_{200m} - M_{500c}$ relations.  That in turn would result in a lower $\sigma_8$ value obtained from the MaxBCG sample than reported in \citet{Rozo2010}, the latter of which is currently in agreement with $\sigma_8$ values obtained from X-ray-selected cluster samples \citep{Henry2009,Vikhlinin2009b, Mantz2010}.

Some other recent papers have added more data and analysis shedding light on the measured discrepancy in SZ signal found by {\em{{\it{Planck}}}}. \citet{Biesiadzinski2012} argue that there is not really a discrepancy if one considers the $2\sigma$ uncertainty on the $N_{200m} - M_{500c}$ relation in \citet{Rozo2009}.  However, if \citet{Rozo2010} used an $N_{200m} - M_{500c}$ relation that had a normalization that is $2\sigma$ away from the true value, then again it is surprising that they find a constraint on $\sigma_8$ with an optical cluster sample that is in such good agreement with the $\sigma_8$ constraints found using X-ray-selected samples.  Such a shift in the $N_{200m} - M_{500c}$ normalization would also put the derived cosmological constraints in some tension with previous results \citep{Rozo2012Pap3}.  \citet{Angulo2012}, using Millennium-XXL simulations, argue along the same lines as \citet{Biesiadzinski2012} suggesting that the normalization of the richness-mass relation is incorrect (see last paragraph of Section 4 in that work). 

\citet{Bauer2012} tested the $N_{200m} - M_{500c}$ relation by verifying the weak-lensing cluster mass estimates upon which this relation was calibrated.  They did this by using gravitational lensing magnification of type I quasars in the backgrounds of these clusters.  Their results support the mass normalization of the \citet{Rozo2009} $N_{200m} - M_{500c}$ relation.  Recent work by \citet{Rozo2012} suggests that part of the {\em{{\it{Planck}}}} measured discrepancy may be due to different X-ray instrument calibrations between {\em{{\it{Chandra}}}}  and {\em{{\it{XMM-Newton}}}} and/or differing analysis procedures when dealing with data from each instrument (see Figure 3 of that work).  \citet{Rozo2012} argue that X-ray scaling relations derived from {\em{{\it{Chandra}}}} observations may be systematically high compared to those from {\em{{\it{XMM-Newton}}}} observations.   Since the $N_{200m} - M_{500c}$ relation was calibrated using an $L_{\rm X} - M_{500c}$ prior derived from {\em{{\it{Chandra}}}} observations \citep{Vikhlinin2009b}, and the \citet{Arnaud2010} profile was derived using {\em{{\it{XMM-Newton}}}} observations, this may be responsible for part of the discrepancy.  \citet{Rozo2012Pap3, Rozo2012Pap2, Rozo2012Pap1} include further discussion in this direction and claim that three ingredients may be responsible for the tension between measured and expected SZ signals measured by {\em{{\it{Planck}}}} for the MaxBCG sample.  These ingredients are a mass-richness relation that is normalized high by $10\%$, {\em{{\it{Chandra}}}} derived X-ray masses from \citet{Vikhlinin2009b} that are biased low by $~20\%$ due to non-thermal pressure, and {\em{{\it{XMM-Newton}}}} X-ray derived masses that are $\sim10$ to $20\%$ lower than those derived from {\em{{\it{Chandra}}}}.  More data will clarify this issue. 

Given the above discussion, the interpretation of both the ACT results presented here and the Planck results stacking MaxBCG clusters remains unclear.  Since each possibility mentioned above, on its own, would be unlikely to explain the full discrepancy between expectation and measurement of both data sets, it is likely that multiple effects are responsible for the low SZ signals.  It is possible that a combination of factors is lowering the SZ signal equally for both ACT and Planck, and that offsets between gas peaks and BCGs result in the remaining discrepancy between ACT and Planck measurements.  We have shown that even in this scenario however, a larger offset distribution than is seen in X-ray-selected cluster samples is still required to explain the discrepancy between ACT and Planck measured signals.

Further investigation is required using multi-wavelength techniques to fully contextualize the SZ signal measured by both {\em{{\it{Planck}}}} and ACT for this sample of optically-selected clusters.  High-resolution SZ instruments such as CARMA and MUSTANG2 as well as more sensitive datasets such as those forthcoming from ACTPol and SPTPol will provide more clarity.  This will yield a better understanding of the astrophysics of galaxy clusters.  Such knowledge will also be vital to assess the use of galaxy clusters as probes of structure growth and the cosmological parameters of the Universe.  
\acknowledgments
NS would like to thank James Bartlett, Sarah Hansen, Jean-Baptiste Melin, Eduardo Rozo, Kendrick Smith and Risa Wechsler for useful discussions as well as Rocco Piffarett and Eli Rykoff for providing the MCXC catalog and the new richness MaxBCG catalog respectively.  NS would also like to thank the organizers of the KITP Monsters Inc. conference, which facilitated further discussions.  

This work was supported by the U.S. National Science Foundation through awards AST-0408698 for the ACT project, and PHY-0355328, AST-0707731 and PIRE-0507768 (award number OISE-0530095). The PIRE program made possible exchanges between Chile, South Africa, Spain and the U.S. that enabled this research program.   Funding was also provided by Princeton University, the University of Pennsylvania, and a Canada Foundation for Innovation (CFI) award to UBC.  Computations were performed on the GPC supercomputer at the SciNet HPC Consortium. SciNet is funded by: the Canada Foundation for Innovation under the auspices of Compute Canada; the Government of Ontario; Ontario Research Fund -- Research Excellence; and the University of Toronto.  ACT is on the Chajnantor Science preserve, which was made possible by the Chilean Comisi\'{o}n Nacional de Investigaci\'{o}n Cient\'{i}fica y Tecnol\'{o}gica (CONICYT).  

NS is supported by the National Science Foundation under Award No. 1102762.  During the completion of this work, NS was also supported by the U.S. Department of Energy contract to SLAC no. DE-AC3-76SF00515 and in part by the National Science Foundation under Grant No. 1066293 and the hospitality of the Aspen Center for Physics.   The data will be made public through LAMBDA (http://lambda.gsfc.nasa.gov/) and the ACT website (http://www.physics.princeton.edu/act/).

\bibliographystyle{hapj}
%\bibliography{refs}

\end{document}